\documentclass[twocolumn,
showpacs,amsmath,amssymb,aps,prb,preprintnumbers]{revtex4-1}
\usepackage{amssymb}

\usepackage{graphicx}
\usepackage{amsmath}
\usepackage{graphicx}
\usepackage{dcolumn}
\usepackage{bm}

\input{tcilatex}

\begin{document}

\preprint{APS}

\title{Bound-magnetic-polaron molecule in diluted magnetic semiconductors}%

\author{Henryk Bednarski}
\email{hbednarski@cmpw-pan.edu.pl}
\author{Jozef Spa\l ek}%
\email{ufspalek@if.uj.edu.pl}
\affiliation{$^{\ast }$Centre of Polymer and Carbon Materials, Polish Academy of Sciences, M. Curie-Sk\l odowska 34, 41-819 Zabrze, Poland \\ $^{\dagger }$Marian Smoluchowski Institute of Physics, Jagiellonian
University, Reymonta 4, 30-059 Krak\'{o}w, Poland}%

\date{\today}%

\begin{abstract}
We formulate a complete microscopic theory of a coupled pair of bound
magnetic polarons, the bound-magnetic-polaron molecule (BMPM) in a diluted
magnetic semiconductor (DMS) by taking into account both a proper two-body
nature of the impurity-electron wave function and within the general
spin-rotation-invariant approach to the electronic states. We also take into
account both the Heisenberg and the antiferromagnetic kinetic-exchange
interactions, as well as the ferromagnetic coupling within the common spin
BMPM cloud. The thermodynamic fluctuations of the spin cloud within the
polaron effective Bohr radius of each polaron are taken as Gaussian.

\pacs{75.10.-b, 75.45.+j, 75.30.Et}%
\end{abstract}

\maketitle

\section{Introduction}

The bound magnetic polarons, single and molecule, represent the charge
carrier quantum states formed when the magnetization fluctuations influence
both their binding and a nontrivial thermal behavior. These states have been
discussed in novel materials encompassing diluted magnetic semiconductors, 
\cite{1} ferromagnetic perovskites,\cite{Teresa} and dilute ferromagnetic
oxides.\cite{Coey} For example, the origin of ferromagnetism in diluted
magnetic semiconductors (DMS) such as Ga$_{1-x}$Mn$_{x}$As, is intensively
discussed in the recent years\cite{1} in view of their potential application
in spintronics. The origin of ferromagnetism\cite{1} poses a very nontrivial
question in view of the dominant role of antiferromagnetic superexchange in
all of those materials when the carrier concentration is very low.\cite{2}
Therefore, it is crucial to describe these interactions accurately in the
physically tractable situation. Here we propose a soluble model of two
interacting impurity electrons forming a bound magnetic-polaron molecule (BMPM).

Magnetic interaction between a single electron located on a shallow impurity
and the localized magnetic moments in diluted magnetic semiconductors (DMS)
has been studied intensively for some time.\cite{3}$^{-}$\cite{9} In the
first period, the influence of classical fluctuations of magnetization on
quantum states of the impurity electron have been analyzed.\cite{3}$^{-}$%
\cite{9}, thus leading to the concept of bound magnetic polaron (BMP). The
main result obtained was to demonstrate that thermodynamic fluctuations
suffice to produce a spontaneous spin splitting of the impurity states. A
renewed interest in the present decade was stimulated by the possibility of
ferromagnetic interpolaron interaction,\cite{10}$^{-}$\cite{12} which would
contribute in a fundamental manner to the origin and properties of
ferromagnetic DMS systems and other materials.\cite{Teresa}$^{-}$\cite{Coey}
In Ref. \onlinecite{10} the authors extended the theory of single BMP\cite{3}%
$^{-}$\cite{9} to the bipolaron case in the limit when the mutual
interaction is represented by hopping of electrons between the two spatially
separated impurities, that leads among others to the antiferromagnetic
kinetic exchange between them.\cite{13} The magnetic cloud is represented
then by an effective magnetic field, which is oriented arbitrarily and
optimized, but its amplitude is a parameter of the approach.\cite{11} In the
later version of the approach,\cite{12} the authors approximate the
two-impurity wave function by piecewise constant values. They consider also
a detailed thermodynamics of the resultant spin model and optimize the
coupling parameters. The explicit solution is discussed in the limit of
large interpolaron distance. Due to complexity of the interacting BMP pair
problem existing theories are still incomplete. For instance, the model
developed by Angelescu and Bhatt\cite{11} of the system of two
nonoverlapping polarons is analyzed via a generalized Hubbard type
Hamiltonian with hopping (matrix element $t$) and Coulomb interaction
(energy $U$) turned on, is thought of as being complementary to the model of
overlapping polaron pair of Ref. \onlinecite{10}. Other important and still
awaiting for solution problems are concerned with the acceptor type
interacting BMP's in II-VI and III-V DMS. Recent achievements in
understanding the ferromagnetism in these classes of materials clearly show
importance of an accurate treating of the non-hydrogeniclike character of
the hole wave function and its influence by the central cell corrections. 
\cite{in1}$^{-}$\cite{in5}

In this paper we introduce and solve BMP molecule Hamiltonian for the case of two donors. It is shown
that our model unifies existing approaches to an interacting polaron pair:
The Wolff-Bhatt-Durst (thereafter referred as to WBD) model\cite{12} of
overlapping polaron pair and the multiple-level generalized Hubbard model of
Angelescu and Bhatt (thereafter referred as to AB) with random fields\cite{11}.
The WBD model is completed by providing whole microscopic justification of
their Hamiltonian parameters. While, the AB model of nonoverlapping,
equal-magnitude and large polarons is generalized to the case of overlapping
and arbitrary-magnitude interacting polaron pair, the extension is limited
here only to the regular Hubbard model (i.e. within one lowest-energy level
on each polaron site only). Furthermore, we solve the resulting Hamiltonian
within the continuum-medium and the effective-mass approximations for the
donor case. In this manner, we extend the BMP model to the microscopic model
of BMP molecule. This constitutes an accurate ground for the future analysis
concerning acceptor-type BMP molecules and the polaron lattices.

The structure of the paper is as follows. In Section II we formulate our
model of BMPM. In Section III we diagonalize the effective Hamiltonian for
the molecule. In Section IV the thermodynamical properties are considered,
whereas Section V contains concluding remarks. The Appendixes A and B
provide details of analytic calculations.

\section{The BMP molecule model}

We start by considering two interacting BMPs in DMS with a random
distribution of localized spins. Within the continuum-medium and the
effective-mass approximations, one can write Hamiltonian in the form: 
\begin{equation*}
H\equiv H_{I}+H_{II}\equiv -J_{c}\left[ \widehat{\mathbf{S}}(\mathbf{r}%
_{1})\cdot \widehat{\mathbf{s}}_{1}+\widehat{\mathbf{S}}(\mathbf{r}%
_{2})\cdot \widehat{\mathbf{s}}_{2}\right] +H_{II}
\end{equation*}
\begin{equation}
=H_{I}-\frac{\hbar ^{2}}{2m^{\ast }}\left( \mathbf{\nabla }_{1}^{2}+\mathbf{%
\nabla }_{2}^{2}\right)  \label{1}
\end{equation}
\begin{equation*}
-\frac{e^{2}}{\varepsilon }\left[ \frac{1}{r_{a1}}+\frac{1}{r_{a2}}+\frac{1}{%
r_{b1}}+\frac{1}{r_{b2}}-\frac{1}{r_{12}}\right] \,,
\end{equation*}
where $m^{\ast }$ is the impurity-electron effective mass, $a$ and $b$ label
the two-impurity sites, $l=1,2$ label the two electrons, $r_{a1}$, $r_{a2}$, 
$r_{b1}$, $r_{b2}$, $r_{12}$, $r_{1}$, $r_{2}$ and $R_{ab}$ are
corresponding relative distances appearing in the problem, $\varepsilon $ is
the static dielectric constant, $J_{c}$ is the exchange integral of the
contact Fermi \ (s-d) interaction between localized spins $\left\{ \widehat{%
\mathbf{S}}_{i}\right\} $ and those of impurity carriers, $\{\widehat{%
\mathbf{s}}_{l}\}.$ $\widehat{\mathbf{S}}(\mathbf{r}_{l})=\sum\nolimits_{i}%
\widehat{\mathbf{S}}_{i}\delta (\mathbf{r}_{l}-\mathbf{R}_{i})$ is the
spin-density operator with the sum running over sites occupied by magnetic
ions (Mn$^{2+}$). $H_{II}$ is the hydrogen-molecule part which supplemented
with the s-d coupling of electrons to localized spins ($H_{I}$).

The solution of BMP-pair problem is complex.\cite{10}$^{-}$\cite{12} From
one side, a systematic approach to the BMP lattices and hydrogen molecules
can be achieved within the second-quantization formalism,\cite{14} but from
the other, the spin part of $H$ is dependent on the positions of the
localized-moment densites. This fact makes the problem more complex to solve.
However, it becomes tractable when one tries to diagonalize $H$ as
consisting of the hydrogeniclike molecule Hamiltonian $H_{II}$ influenced by
the perturbation $H_{I}.$ This suggests to approach the solution of BMPM by
selecting as trial states the eigenstates of the set of mutually compatible
observables $\{\mathbf{s}_{tot}^{2},s_{tot}^{z}\}$, where $\mathbf{s}_{tot}$
is the total spin of the BMPM. In fact, as mobile carriers interact via an
isotropic Coulomb repulsion, only rotations of the whole molecule conserve
their relative distance, thus leaving their energy unchanged. Moreover, we
know that in the Heitler-London approximation the Hamiltonian commutes with
both $\mathbf{s}_{tot}^{2}$ and $s_{tot}^{z},$ it is thus diagonal with
three-fold degenerate triplet states. Such an approach greatly simplifies the
problem, as in practice we have solutions for $H_{II}$ within the eigenstates of 
$\{\mathbf{s}_{tot}^{2},s_{tot}^{z}\}.$ Here we consider more general than Heitler-London
solution of the hydrogen molecule by including also ionic configurations. In
this case, $H_{II}$ is not diagonal in the basis formed from eigenstates of $%
\mathbf{s}_{tot}^{2}$ and $s_{tot}^{z}.$ Fortunately, for the off-diagonal
elements, the hopping couples only the different singlet states.

Next, we express $H_{I}$ in terms of the creation and annihilation operators.
Details of calculations of the matrix representation of $H_{I}$ within the
first quantization formalism are presented in Appendix A. Our methodology
bases on an explicit demonstration that the postulated occupation-number
representation of the polaronic part of the BMPM Hamiltonian leads to the same
expectation values as those obtained within the first quantization scheme.
Thus, we assume $H_{I}$ in the following form: 
\begin{equation}
H_{I}=\frac{1}{2}a_{as}^{\dagger }(\mathbf{\Delta }_{a}\cdot \mathbf{\sigma }%
)_{ss^{\prime }}a_{as^{\prime }}+\frac{1}{2}a_{bs}^{\dagger }(\mathbf{\Delta 
}_{b}\cdot \mathbf{\sigma })_{ss^{\prime }}a_{bs^{\prime }}^{\prime },
\label{2}
\end{equation}
where $a_{cs}^{\dagger }$ ($a_{cs}$) is the creation (annihilation) operator
for the state on impurity $c=a,b$ with the spin $s$ $(s=\uparrow ,\downarrow
)$ and the polaron exchange fields $\mathbf{\Delta }_{c}$ are defined as: 
\begin{equation}
\mathbf{\Delta }_{c}\equiv \frac{\alpha }{g\mu _{B}}\int w_{c}^{\ast }(%
\mathbf{r})\mathbf{M}(\mathbf{r})w_{c}(\mathbf{r})d^{3}r,\quad c=a,\ b,
\label{3}
\end{equation}
where $\mathbf{M}(\mathbf{r})=-g\mu _{B}N_{0}x\left\langle \chi _{S}\left| 
\widehat{\mathbf{S}}(\mathbf{r})\right| \chi _{S}\right\rangle $ is the
local magnetization per unit volume; $N_{0}=n_{0}/v_{0}$ is number of atoms
per unit volume containing fraction $x$ of magnetic atoms, $w_{a}(\mathbf{r}%
) $ and $w_{b}(\mathbf{r})$ are the orthogonal molecular wave functions: 
\begin{equation}
w_{a,b}(r)=\beta \lbrack \psi _{a,b}(r)-\gamma \psi _{b,a}(r)],  \label{4}
\end{equation}
which are build from the single-particle atomic wave functions $\psi _{a}$
and $\psi _{b},$ being the solution of the corresponding single-particle
hydrogeniclike Schr\"{o}dinger equation with the effective Bohr radius a$%
_{B}.$ $\beta $\ and $\gamma $ are the mixing coefficients: 
\begin{equation}
\beta =\frac{1}{\sqrt{2}}\left[ \frac{1}{1-S^{2}}+\frac{1}{\sqrt{1-S^{2}}}%
\right] ^{1/2},\quad \gamma =\frac{S}{1+\sqrt{1-S^{2}}},  \label{5}
\end{equation}
determined by imposing the ortogonality and the normalization conditions,
whereas $S=\left\langle \psi _{a}(r)|\psi _{b}(r)\right\rangle $ is the
overlap integral. It is important to note here, that when calculating $%
\mathbf{\Delta }_{c}$ we replaced the spin density operator $\widehat{%
\mathbf{S}}(\mathbf{r})$ by its quantum mechanical average, thereby
introducing the adiabatic and mean field approximations. This however is
acceptable here, because the same approximations are assumed in the single
BMP theory and provide good results.\cite{3} The BMP's exchange fields $%
\mathbf{\Delta }_{a}$ and $\mathbf{\Delta }_{b}$ may be oriented in an
arbitrary direction and we do not make any a priori restriction for their
magnitude. In our model those polaronic exchange fields are overlapping,
through the presence of the mixing coefficients in the definition of $%
\mathbf{\Delta }_{a}$ and $\mathbf{\Delta }_{b}$. Next, we introduce the six
eigenstates of the carriers total spin operator $\widehat{\mathbf{s}}=%
\widehat{\mathbf{s}}_{1}+\widehat{\mathbf{s}}_{2}$: three triplet ($s=1$)
and three singlet ($s=0$) states.\cite{14}\ The triplet states are: 
\begin{equation}
\left| 1\right\rangle =\frac{1}{\sqrt{2}}(a_{a\uparrow }^{\dagger
}a_{b\downarrow }^{\dagger }+a_{a\downarrow }^{\dagger }a_{b\uparrow
}^{\dagger })\left| 0\right\rangle ,  \label{6}
\end{equation}
\begin{equation}
\left| 2\right\rangle =a_{a\uparrow }^{\dagger }a_{b\uparrow }^{\dagger
}\left| 0\right\rangle ,\quad \left| 3\right\rangle =a_{a\downarrow
}^{\dagger }a_{b\downarrow }^{\dagger }\left| 0\right\rangle ,  \label{6a}
\end{equation}
and the corresponding three singlet states are: 
\begin{equation}
\left| 4\right\rangle =\frac{1}{\sqrt{2}}(a_{a\uparrow }^{\dagger
}a_{b\downarrow }^{\dagger }-a_{a\downarrow }^{\dagger }a_{b\uparrow
}^{\dagger })\left| 0\right\rangle ,  \label{7}
\end{equation}
\begin{equation}
\left| 5\right\rangle =\frac{1}{\sqrt{2}}(a_{a\uparrow }^{\dagger
}a_{a\downarrow }^{\dagger }+a_{b\downarrow }^{\dagger }a_{b\uparrow
}^{\dagger })\left| 0\right\rangle ,  \label{7a}
\end{equation}
and 
\begin{equation}
\left| 6\right\rangle =\frac{1}{\sqrt{2}}(a_{a\uparrow }^{\dagger
}a_{a\downarrow }^{\dagger }-a_{b\downarrow }^{\dagger }a_{b\uparrow
}^{\dagger })\left| 0\right\rangle .  \label{7b}
\end{equation}
It is also more convenient to transform the BMP exchange fields $\mathbf{%
\Delta }_{a}$ and $\mathbf{\Delta }_{b}$ to $\mathbf{\Delta }^{+}$ and $%
\mathbf{\Delta }^{-}$ according to following prescriptions: 
\begin{equation}
\mathbf{\Delta }^{+}\equiv \frac{1}{2}(\mathbf{\Delta }_{a}+\mathbf{\Delta }%
_{b})\quad \text{and }\mathbf{\Delta }^{-}\equiv \frac{1}{2}(\mathbf{\Delta }%
_{a}-\mathbf{\Delta }_{b}),  \label{8}
\end{equation}
which enable us to write the effective polaron fields as: 
\begin{equation}
\mathbf{\Delta }^{\pm }[\mathbf{M}]=\frac{\alpha }{g\mu _{B}}\int
d^{3}r[\left| w_{a}(r)\right| ^{2}\pm \left| w_{b}(r)\right| ^{2}]\mathbf{M}%
(r).  \label{8a}
\end{equation}
This in turn leads to $H_{I}$ in the form: 
\begin{equation*}
H_{I}=\frac{1}{2}\sum_{ss^{\prime }}(a_{as}^{\dagger }(\mathbf{\Delta }%
^{+}\cdot \mathbf{\sigma })_{ss^{\prime }}a_{as^{\prime }}+a_{bs}^{\dagger }(%
\mathbf{\Delta }^{+}\cdot \mathbf{\sigma })_{ss^{\prime }}a_{bs^{\prime
}}^{\prime }
\end{equation*}
\begin{equation}
+a_{as}^{\dagger }(\mathbf{\Delta }^{-}\cdot \mathbf{\sigma })_{ss^{\prime
}}a_{as^{\prime }}-a_{bs}^{\dagger }(\mathbf{\Delta }^{-}\cdot \mathbf{%
\sigma })_{ss^{\prime }}a_{bs^{\prime }}^{\prime }).  \label{e9}
\end{equation}
Now, we select the direction of the global quantization axis as aligned with 
$\mathbf{\Delta }^{-}$. This choice leads to the following $6\times 6$
singlet-triplet matrix representation of $H_{I}$: 
\begin{equation}
\left( 
\begin{array}{cccccc}
0 & \Delta ^{+}\frac{\sin \theta }{\sqrt{2}}e^{i\varphi } & \Delta ^{+}\frac{%
\sin \theta }{\sqrt{2}}e^{-i\varphi } & \Delta ^{-} & 0 & 0 \\ 
\Delta ^{+}\frac{\sin \theta }{\sqrt{2}}e^{-i\varphi } & \Delta ^{+}\cos
\theta & 0 & 0 & 0 & 0 \\ 
\Delta ^{+}\frac{\sin \theta }{\sqrt{2}}e^{i\varphi } & 0 & -\Delta ^{+}\cos
\theta & 0 & 0 & 0 \\ 
\Delta ^{-} & 0 & 0 & 0 & 0 & 0 \\ 
0 & 0 & 0 & 0 & 0 & \Delta ^{-} \\ 
0 & 0 & 0 & 0 & \Delta ^{-} & 0
\end{array}
\right)  \label{10}
\end{equation}
where $\theta $ and $\varphi $ are respectively, the azimuthal and the polar
angles between the exchange field $\mathbf{\Delta }^{+}$ and $\mathbf{\Delta 
}^{-}.$

For the solution of the BMPM problem, the polaronic part $H_{I}$ must be
completed with the hydrogenlike molecule part $H_{II},$ written also in the
second quantization form. Such representation of $H_{II}$ is known\cite{13}$%
^{-}$\cite{14} so, we can write it directly as: 
\begin{multline}
H_{II}=\epsilon _{a}n_{a}+\epsilon _{b}n_{b}+t\sum_{s}\left( a_{as}^{\dagger
}a_{as}+a_{bs}^{\dagger }a_{bs}\right) +U_{a}n_{a\uparrow }n_{a\downarrow }
\\
+U_{b}n_{b\uparrow }n_{b\downarrow }-2J\widehat{\mathbf{s}}_{1}\cdot 
\widehat{\mathbf{s}}_{2}+\left( K+\frac{1}{2}J\right) n_{a}n_{b} \\
+J(a_{a\uparrow }^{\dagger }a_{a\downarrow }^{\dagger }a_{b\downarrow
}a_{b\uparrow }+H.c.)+V\sum_{s}[(n_{as}+n_{as}) \\
\times (\widehat{a}_{a\overline{s}}^{\dagger }a_{b\overline{s}}+\widehat{a}%
_{b\overline{s}}^{\dagger }a_{a\overline{s}})],
\end{multline}
where, $\epsilon _{c}$ is the atomic level position, $t$ the hopping between
the impurity states and $U_{c}$ the intra-atomic Coulomb interaction. The
remaining terms represent respectively, the interatomic (Heisenberg)
exchange, the pair hopping, and the so-called correlated hopping. In the
singlet-triplet basis, $H_{II}$ has relatively simple form\cite{14}, so we
may write: 
\begin{equation}
H_{II}=\left( 
\begin{array}{cccccc}
L & 0 & 0 & 0 & 0 & 0 \\ 
0 & L & 0 & 0 & 0 & 0 \\ 
0 & 0 & L & 0 & 0 & 0 \\ 
0 & 0 & 0 & L+2j & T & 0 \\ 
0 & 0 & 0 & T & L_{6}+2j & 0 \\ 
0 & 0 & 0 & 0 & 0 & L_{6}
\end{array}
\right) \,,  \label{12}
\end{equation}
where: 
\begin{equation}
L=2\epsilon +K-j,\,L_{6}=2\epsilon +U-j,\text{ and}\,T=2(t+V).  \label{13}
\end{equation}
For identical impurities both the atomic level positions $\epsilon _{a}$ and $%
\epsilon _{b}$ and the intra-atomic Coulomb interactions $U_{a}$ and $U_{b}$
are equal (i.e. $\epsilon _{a}=$ $\epsilon _{b}\equiv \epsilon $ and $U_{a}=$
$U_{b}=U$). We note here, that $H_{II}$ is solved in the local coordinate
system, with $z$ axis aligned with the direction of the molecular bond, and
then Hamiltonian $H_{II}$ is reoriented towards the global quantization
axis. This last step is quite trivial due to the rotational invariance of $%
H_{II}$.

Having determined the BMP molecule Hamiltonian, we discuss next its relation
to the with existing models of Angelescu-Bhatt (AB) and Wolff--Bhatt-Durst
(WBD).\cite{10}$^{-}$\cite{12} These earlier models of interacting BMP pair,
while being significant theoretical achievements, are subject to certain
important limitations, clearly stated by their authors. Here, for a purpose
of completeness, we enumerate some of them. As in our model each site
contributes a single s-type orbital, the AB model appears as more advanced,
because it accounts for excited-states transitions from 1s level up to 3d
level. Nevertheless, it should be kept in mind, that AB model treats BMPs as
nonoverlapping, equal and large field within the perturbatoion approach with
respect to their mutual interaction, hence neglecting important processes
even in the simplest one-level situation. For instance, AB model does not
predict ferromagnetic ground state in that case. Their assumption about
equal magnitudes of BMPs exchange fields creates a serious problem with
treating them as thermodynamically fluctuating quantities. From the other
side, trusting in predictions from WBD model is weakened by not fully
microscopic derivation of their Hamiltionian, including their parameters.

\begin{figure}[tbp]
\includegraphics{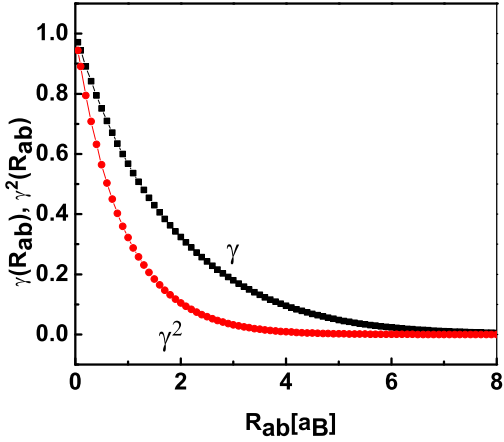} \label{fig1}
\caption{(Color online) Mixing coefficients $\protect\gamma $ and $\protect%
\gamma ^{2}$ as a function of interimpurity distance $R_{ab}$ expressed in
the units of $a_{B}.$}
\end{figure}
Therefore, it is important to recognize consequences of such approximations,
even in a relatively simplest case. In this work the model is formulated as applicable
beyond those limits namely, our BMPs are overlapping, have arbitrary
magnitude of the exchange fields and their mutual interaction is accounted
for within molecular formalism. Moreover, we are able to consider also
thermodynamics, including thermal fluctuations of BMPs exchange
fields.

\begin{figure}[tbp]
\includegraphics{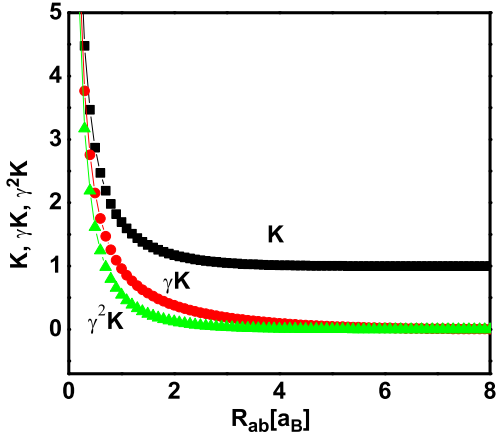} \label{fig2}
\caption{(Color online) $K,$ $\protect\gamma K$ and $\protect\gamma ^{2}K$ \
(in units of $\protect\alpha $) as a function of interimpurity distance $%
R_{ab}$ expressed in the units of $a_{B}.$}
\end{figure}

To demonstrate the generality of our approach more clearly, we map polaronic
part $H_{I}$ into corresponding parts of the WBD model Hamiltonian (the
whole WBD model contains also the antiferromagnetic Heisenberg exchange
interaction between carrier spins, which we do not specify explicitly).
Namely, we can rewrite $H_{I}$ in the spin operator form: 
\begin{equation}
H_{I}=\mathbf{\Delta }_{a}\cdot \widehat{\mathbf{s}}_{1}+\mathbf{\Delta }%
_{b}\cdot \widehat{\mathbf{s}}_{2},  \label{14}
\end{equation}
and take into account the explicit form of the exchange fields $\mathbf{%
\Delta }_{a}$ and $\mathbf{\Delta }_{b}$ which yields: 
\begin{align}
H_{I}& =\frac{\alpha }{g\mu _{B}}\beta ^{2}[\widehat{\mathbf{s}}_{1}\cdot
\int \left| \psi _{a}\right| ^{2}\mathbf{M}d^{3}r+\widehat{\mathbf{s}}%
_{2}\cdot \int \left| \psi _{b}\right| ^{2}\mathbf{M}d^{3}r]  \label{15} \\
& -\frac{\alpha }{g\mu _{B}}\beta ^{2}\gamma (\widehat{\mathbf{s}}_{1}+%
\widehat{\mathbf{s}}_{2})\cdot \int (\psi _{a}^{\ast }\psi _{b}+\psi
_{b}^{\ast }\psi _{a})\mathbf{M}d^{3}r  \notag \\
& +\frac{\alpha }{g\mu _{B}}\beta ^{2}\gamma ^{2}[\widehat{\mathbf{s}}%
_{1}\cdot \int \left| \psi _{b}\right| ^{2}\mathbf{M}d^{3}r+\widehat{\mathbf{%
s}}_{2}\cdot \int \left| \psi _{a}\right| ^{2}\mathbf{M})d^{3}r].  \notag
\end{align}
This expression should be compared with following polaronic part of the WBD
model Hamiltonian:\cite{12} 
\begin{equation}
H_{WBD}=K[\widehat{s}_{1}\cdot \mathbf{S}_{1}+\widehat{s}_{2}\cdot \mathbf{S}%
_{2}]+K^{\prime }(\widehat{s}_{1}+\widehat{s}_{2})\cdot \mathbf{S}_{3},
\label{16}
\end{equation}
where now $K$ and $K^{\prime }$ are the WBD model parameters and $\mathbf{S}%
_{3}$ is the total spin of magnetic ions in overlapping region. A direct
mapping can be established by introducing the following definitions: 
\begin{multline}
K\equiv \alpha \beta ^{2},\ K^{\prime }\equiv -\gamma K,\quad \ \mathbf{S}%
_{1(2)}\equiv \int \left| \psi _{a(b)}\right| ^{2}\frac{\mathbf{M}}{g\mu _{B}%
}d^{3}r  \\
\text{and\quad }\mathbf{S}_{3}\equiv \int (\psi _{a}^{\ast }\psi _{b}+\psi
_{b}^{\ast }\psi _{a})\frac{\mathbf{M}}{g\mu _{B}}d^{3}r,  \label{h19}
\end{multline}
which allow to rewrite $H_{I}$ into the final form: 
\begin{multline}
H_{I}=K[\widehat{\mathbf{s}}_{1}\cdot \mathbf{S}_{1}+\widehat{\mathbf{s}}%
_{2}\cdot \mathbf{S}_{2}]+K^{\prime }(\widehat{\mathbf{s}}_{1}+\widehat{%
\mathbf{s}}_{2})\cdot \mathbf{S}_{3}  \\
+\gamma ^{2}K[\widehat{\mathbf{s}}_{1}\cdot \mathbf{S}_{2}+\widehat{\mathbf{s%
}}_{2}\cdot \mathbf{S}_{1}].  \label{h20}
\end{multline}
In Fig. 1 we have plotted $\gamma $ and $\gamma ^{2}$ as a function of
interpolaron distance $R_{ab}$, where in Fig. 2 we show the same radial
dependence for $K$, $\gamma K$ and $\gamma ^{2}K.$ As can be seen from these
plots only $K$ is different from zero at large distances and in this limit
is equal to that corresponding to isolated BMPs. In $H_{I},$ the parameters $%
\gamma $ and $\gamma ^{2}$ appear as interaction couplings and the
corresponding three terms describe contribution from the zero-, first- and
second-order processes, respectively. One can also see, that at large
interpolaron distances, for which $\gamma ^{2}\rightarrow 0,$ $H_{I}$ can be
reduced to the bipolaron part of WBD model Hamiltonian. In this manner, we
have provided the microscopic derivation of the WBD model and the
microscopic meaning of their parameters. Also, the derived relation of BMP
exchange fields $\mathbf{\Delta }_{a}$ and $\mathbf{\Delta }_{b}$ to the
carrier ortognalized single particle wave functions shows that the magnitude
of these quantities depends on the interpolaron distance $R_{ab},$ the
feature which is completely neglected in the AB model.

\section{Solution of the BMP molecule Hamiltonian}

Diagonalization of the BMPM Hamiltonian requires the solution of the
eigenequation $HV=EV,$ with $H$ given by:%
\begin{widetext}
\begin{equation}
H=\left( 
\begin{array}{cccccc}
L & \Delta ^{+}\frac{\sin \theta }{\sqrt{2}}e^{i\varphi } & \Delta ^{+}\frac{\sin \theta }{\sqrt{2}}e^{-i\varphi } & \Delta ^{-} & 0 & 0 \\ 
\Delta ^{+}\frac{\sin \theta }{\sqrt{2}}e^{-i\varphi } & L+\Delta ^{+}\cos
\theta  & 0 & 0 & 0 & 0 \\ 
\Delta ^{+}\frac{\sin \theta }{\sqrt{2}}e^{i\varphi } & 0 & L-\Delta
^{+}\cos \theta  & 0 & 0 & 0 \\ 
\Delta ^{-} & 0 & 0 & L+2j & T & 0 \\ 
0 & 0 & 0 & T & L_{6}+2j & \Delta ^{-} \\ 
0 & 0 & 0 & 0 & \Delta ^{-} & L_{6}
\end{array}
\right) .  \label{a1}
\end{equation}\end{widetext}%
%
Its form does not allow for an exact analytical diagonalization. Therefore,
before solving it numerically, we discuss first some limiting situations,
for which approximate analytical solutions can be obtained.

\subsection{Saturation limit}

We first solve BMPM effective Hamiltonian for pair of donors in the
saturation limit, i.e. with $\mathbf{M}(\mathbf{r})=\mathbf{M}_{sat}$. In
this case, the solution is simplified greatly within our approach, because
we can write: 
\begin{equation}
\Delta _{+}=\frac{\alpha }{g\mu _{B}}M_{sat}\ \mathbf{=\ }5/2\alpha N\quad 
\text{and\quad }\mathbf{\Delta }_{-}=0,  \label{19}
\end{equation}
where $N$ is the number of Mn ions within BMPM which contribute to
magnetization. Subsequent diagonalization of $H$ leads to following exact
solution for the enigenvalues $E_{i}$, $i=1,...,6$: 
\begin{equation}
\left( 
\begin{array}{c}
L \\ 
L+5/2\alpha N \\ 
L-5/2\alpha N \\ 
\frac{1}{2}[(L+L_{6})+4j+\sqrt{\left( L-L_{6}\right) ^{2}+4T^{2}}] \\ 
\frac{1}{2}[(L+L_{6})+4j-\sqrt{\left( L-L_{6}\right) ^{2}+4T^{2}}] \\ 
L_{6}
\end{array}
\right) .  \label{20}
\end{equation}
The eigenvalues $E_{4,5}$ contain antiferromagnetic kinetic exchange
interaction\cite{14} in an explicit form, that competes with the direct
Heisenberg exchange. Nature of the ground state depend on the sign of the
expression\ $\Delta E\equiv E_{5}(\mathbf{\Delta }_{-})-E_{3}(\Delta _{+}),$
which determines the dominant exchange interaction aligning individual
polaron polarization clouds;for positive value the ground state is
ferromagnetic. Note that states belonging to the eigenvalus $E_{1},$ $E_{2}$%
\ and $E_{3}$ are the triplet states. 
\begin{figure}[tbp]
\includegraphics{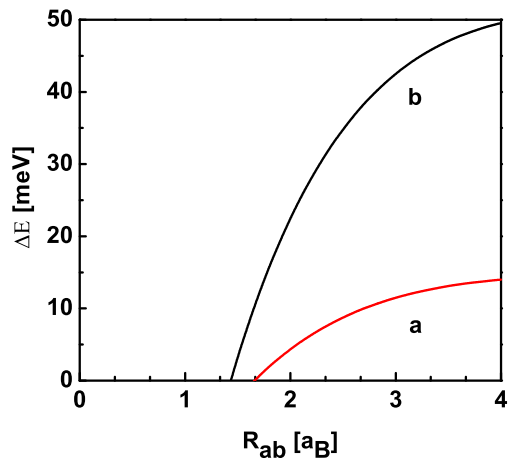} \label{fig3}
\caption{(Color online) Exchange splitting $\Delta E$ dependence on
inter-impurity distance $R_{ab}$ for BMPM with effective concentration $%
\overline{x}=0.027$ of magnetic ions. Lines $a$ and $b$ are calculated with
material parameters corresponding to Cd$_{1-x}$Mn$_{x}$Se and Ga$_{1-x}$Mn$%
_{x}$As$,$ respectively, within the s-state approximation for the impurity
single-particle wave function.}
\end{figure}
We plot in Fig. 3 the singlet-triplet splitting $\Delta E$ as a function of
inter-impurity distance $R_{ab},$ calculated for the material parameters
corresponding to Cd$_{0.95}$Mn$_{0.01}$Se; $N_{0}\alpha =0.28$ eV (line a)$.$%
\cite{3} This naturally is only a formal dependence, which shows a potential
strength of a force stabilizing ferromagnetic state. One sees also that if
the two impurities are to close each other, then nonmagnetic
hydrogen-molecule spin-singlet ground state becomes dominant. In s-type
II-VI DMS, the magnetic susceptibility $\chi $ of localized magnetic ions at
low temperatures takes the form of the Curie-Weiss law $\chi
=C_{M}/(T+T_{0}),$ with $T_{0}>0$.\cite{2} Therefore, the magnetization is
certainly not saturated and considered case appears as describing a
non-realistic situation. The situation changes for p-type ferromagnetic
III-V DMS in which magnetization saturates and presented here model provides
a quantitative argument that for certain range of interpolaron distances the
interactions between BMPs can stabilize ferromagnetism. In Fig. 3. the line
(b) was calculated with values of parameters corresponding to p-d exchange $%
N_{0}\beta =-1.2$ eV confirming our last sentence. Nevertheless, it should
be stressed that the nature of holes wave function in zinc-blende
semiconductors is much more complicated, then \ the used here simple s-type
donor wave function. In other words, the acceptor BMPM is not quite correct
theory.\cite{16}$^{-}$\cite{17} We postpone this discussion to a separate
publication.

\subsection{Large inter-polaron distance limit}

It can be derived by making the transformation $R^{-1}HR$ on the whole $6$x$%
6 $ Hamiltonian matrix, with $R$ being the rotational matrix which
diagonalizes $H_{II}.$\ In explicit form $R$ is the sparse matrix, with all
elements equal zero, except for $R_{44}=R_{55}=\cos (\delta )$, $%
R_{45}=-\sin (\delta )$ and $R_{54}=\sin (\delta ),$ and the rotation angle $%
\delta $ is defined through: 
\begin{equation}
\tan (2\delta )=\frac{4(t+V)}{K-U}.  \label{eqt:2}
\end{equation}
This leads to $H$ in the form:

\begin{widetext}
\begin{equation}
H=\left( 
\begin{array}{cccccc}
L & \frac{2\sin (\theta )}{\sqrt{2}}e^{i\varphi }\Delta ^{+} & \frac{2\sin
(\theta )}{\sqrt{2}}e^{-i\varphi }\Delta ^{+} & \Delta ^{-}\cos (\delta ) & 
-\Delta ^{-}\sin (\delta ) & 0 \\ 
\frac{2\sin (\theta )}{\sqrt{2}}e^{-i\varphi }\Delta ^{+} & L+\Delta
^{+}\cos (\theta ) & 0 & 0 & 0 & 0 \\ 
\frac{2\sin (\theta )}{\sqrt{2}}e^{i\varphi }\Delta ^{+} & 0 & L-\Delta
^{+}\cos (\theta ) & 0 & 0 & 0 \\ 
\Delta ^{-}\cos (\delta ) & 0 & 0 & L_{4} & 0 & \Delta ^{-}\sin (\delta ) \\ 
-\Delta ^{-}\sin (\delta ) & 0 & 0 & 0 & L_{5} & \Delta ^{-}\cos (\delta )
\\ 
0 & 0 & 0 & \Delta ^{-}\sin (\delta ) & \Delta ^{-}\cos (\delta ) & L_{6}
\end{array}
\right) ,  \label{eqt:3}
\end{equation}\end{widetext}%
%
where the diagonal elements $L_{4}$ and $L_{5}$ are: 
\begin{equation}
L_{4,5}\equiv 2\epsilon _{a}+1/2(K+U)+j\pm 1/2[(U-K)^{2}+16(t+V)^{2}]^{1/2}.
\label{a2}
\end{equation}
In the considered here limit of large interpolaron distances one can neglect
the terms proportional to $\sin (\delta ).$ In practice, this limits us to
the region with $R_{ab}\gtrsim 3.5$ a$_{B}.$ In that limit, we can write the
resulting sixth-order eigenequation as: 
\begin{multline}
(E^{2}-E(L_{5}+L_{6})+L_{5}L_{6}-\mathbf{\Delta }^{-2})  \label{30} \\
\{[(E-L)^{2}-\mathbf{\Delta }^{+2}](E^{2}-E(L+L_{4})+LL_{4}-\mathbf{\Delta }%
^{-2}) \\
-(\mathbf{\Delta }^{+}\times \mathbf{\Delta }^{-})^{2}\}=0.
\end{multline}
The eigenvalues $E_{5}$ and $E_{6}$ can be easily determined, but presence
of the vector product $(\mathbf{\Delta }^{+}\times \mathbf{\Delta }^{-})^{2}$
still complicates calculations of remaining four eigenvalues in simple
terms. Formally, this equation can be solved exactly, as in a principle all
roots of the four order equation can be found analytically. However,
solutions will have complicated analytical form, which precludes further
explicit analysis. To overcome this difficulty we regard the expression in $%
\{...\}$ as a function of $E$, say $F(E$), and write: 
\begin{equation}
F(E,\theta )=G(E)+A(\theta ),  \label{31}
\end{equation}
where 
\begin{equation}
A(\theta )=-(\mathbf{\Delta }^{+})^{2}(\mathbf{\Delta }^{-})^{2}\sin
^{2}(\theta ),  \label{32}
\end{equation}
the definition of the function $G(E)$ is self-explanatory. Now, it is easy
to observe that the quantity $A(\theta )$, which is independent of the
energy $E$, influences $F(E,\theta )$ through a downward shift of $G(E).$
Thus we need to analyze properly $G(E)$ and at least take into account
approximately the presence of $A(\theta ).$ Therefore, we can expand $%
F(E,\theta )$ in terms of Taylor series for each eigenvalue separately
around the zeros of $G(E)$: 
\begin{multline}
F(E,\theta )=A(\theta )+\frac{1}{1!}\frac{\partial G(E)}{\partial E}_{\mid
_{E=E_{0i}}}(E-E_{0i})  \label{33} \\
+\frac{1}{2!}\frac{\partial ^{2}G(E)}{\partial E^{2}}_{\mid
_{E=E_{0i}}}(E-E_{0i})^{2}+...=0.
\end{multline}
We solve those equations to the first order and obtain: 
\begin{equation}
E_{i}=E_{0i}-\frac{A(\theta )}{\frac{1}{1!}\frac{\partial G(E,\xi )}{%
\partial E}_{\mid _{E=E_{0i}}}},\quad \text{for }i=1,...,4,  \label{34}
\end{equation}
In effect, we find the following six zero-order eigenvalues $E_{0i}$: 
\begin{equation}
\left( 
\begin{array}{c}
1/2(L+L_{4}-\sqrt{4(\mathbf{\Delta }^{-})^{2}+(L-L_{4})^{2}}) \\ 
L+\Delta ^{+} \\ 
L-\Delta ^{+} \\ 
1/2(L+L_{4}+\sqrt{4(\mathbf{\Delta }^{-})^{2}+(L-L_{4})^{2}}) \\ 
1/2(L_{6}+L_{5}-\sqrt{4(\mathbf{\Delta }^{-})^{2}+(L_{6}-L_{5})^{2}}) \\ 
1/2(L_{6}+L_{5}+\sqrt{4(\mathbf{\Delta }^{-})^{2}+(L_{6}-L_{5})^{2}})
\end{array}
\right)  \label{35}
\end{equation}
Analysis of this expression shows that the nature of the ground state is
determined by the sign of $\Delta E=E_{5}-E_{3}$ expressing the difference
between parallel and antiparallel configuration of the exchange fields.
Moreover, for $\mathbf{\Delta }^{-}=0,$ the eigevalues have the same
functional form as in the exact solution, except that now $\Delta ^{+}$is
not calculated for saturated situation.

\subsubsection{Asymptotic solution for $R_{ab}\longrightarrow \infty $}

In the limit $R_{ab}\longrightarrow \infty ,$ we can neglect the terms
describing interactions between BMPs and two ionic configurations.
Naturally, in resulting expression $\mathbf{\Delta }^{+}$ and $\mathbf{%
\Delta }^{-}$ cannot be neglected. Next, we diagonalize $H_{R_{ab%
\longrightarrow \infty }}$ i.e. write: 
\begin{equation}
H_{R_{ab\longrightarrow \infty }}\widetilde{V}=E\widetilde{V}.  \label{36}
\end{equation}
This equation leads to the related fourth-order equation of the form: 
\begin{multline}
\lbrack (2E_{\infty }-E)^{2}-(\mathbf{\Delta }^{+})^{2}][((2E_{\infty
}-E)^{2}-(\mathbf{\Delta }^{-})^{2}]  \label{37} \\
-(\mathbf{\Delta }^{+})^{2}(\mathbf{\Delta }^{-})^{2}\sin ^{2}(\theta )=0,
\end{multline}
where $E_{\infty }$ $=\epsilon _{a}=\epsilon _{b}.$ The corresponding four
eigenvalues are: 
\begin{widetext}
\begin{equation}
\left( 
\begin{array}{c}
2E_{\infty }+\frac{1}{4}\sqrt{2(\Delta _{+}^{2}+\Delta _{-}^{2})+2\sqrt{(\Delta _{+}^{2}+\Delta _{-}^{2})^{2}-4\Delta _{+}^{2}\Delta _{-}^{2}\cos
^{2}(\theta )}} \\ 
2E_{\infty }-\frac{1}{4}\sqrt{\left[ 2(\Delta _{+}^{2}+\Delta _{-}^{2})+2\sqrt{(\Delta _{+}^{2}+\Delta _{-}^{2})^{2}-4\Delta _{+}^{2}\Delta
_{-}^{2}\cos ^{2}(\theta )}\right] } \\ 
2E_{\infty }+\frac{1}{4}\sqrt{\left[ 2(\Delta _{+}^{2}+\Delta _{-}^{2})-2\sqrt{(\Delta _{+}^{2}+\Delta _{-}^{2})^{2}-4\Delta _{+}^{2}\Delta
_{-}^{2}\cos ^{2}(\theta )}\right] } \\ 
2E_{\infty }-\frac{1}{4}\sqrt{\left[ 2(\Delta _{+}^{2}+\Delta _{-}^{2})-2\sqrt{(\Delta _{+}^{2}+\Delta _{-}^{2})^{2}-4\Delta _{+}^{2}\Delta
_{-}^{2}\cos ^{2}(\theta )}\right] }\end{array}\right)   \label{38}
\end{equation}\end{widetext}%
%
Our asymptotic solutions still depend formally on the angle $\theta $
between $\mathbf{\Delta }^{+}$and $\mathbf{\Delta }^{-}$, whereas the
corresponding solutions for the two isolated BMPs: 
\begin{equation}
\left( 
\begin{array}{c}
2E_{\infty }-\frac{1}{2}(\Delta _{a}+\Delta _{b}) \\ 
2E_{\infty }-\frac{1}{2}(\Delta _{a}-\Delta _{b}) \\ 
2E_{\infty }+\frac{1}{2}(\Delta _{a}-\Delta _{b}) \\ 
2E_{\infty }+\frac{1}{2}(\Delta _{a}+\Delta _{b})
\end{array}
\right)  \label{38a}
\end{equation}
are clearly free of such angular dependence, what reflects an uncorrelated
character of the spatial orientation of the polaron fields. In Appendix B we
prove the equivalence of our solutions given by Eq. (37), with that for the
two isolated polarons. Note, that for finite interpolaron distances, the
limit $\Delta \rightarrow 0$ leads also to a correct solution of the
hydrogeniclike molecule.

\section{Thermodynamics}

Thermodynamic fluctuations of magnetization may strongly influence behavior
of the system. Having determined the eigenvalues, we can construct the free
energy of the BMP pair. To determine these properties we extend the previous
approach\cite{3} devised for a single BMP. For the case of single BMP, the
thermodynamics has been derived by including the contribution coming from
localized magnetic moments starting form the Ginzburg-Landau Hamiltonian: 
\begin{equation}
H_{S}[\mathbf{M}]=\int d^{3}r(\frac{1}{2}\kappa \sum_{j=1}^{3}\left| \mathbf{%
\nabla }M_{j}(\mathbf{r})\right| ^{2}+\frac{1}{2\chi }\mathbf{M}(\mathbf{r}%
)^{2}),  \label{eq:12}
\end{equation}
where the two phenomenological parameters, $\kappa $ and $\chi ^{-1},$ are
the exchange stiffness constant and the inverse static susceptibility,
respectively. For this case, the probability distribution of the exchange
field has been found in the form: 
\begin{equation}
P(\mathbf{\Delta })=\int D\mathbf{M}(\mathbf{r})P[\mathbf{M}(\mathbf{r}%
)]\delta (\mathbf{\Delta }-\mathbf{\Delta }[\mathbf{M}(\mathbf{r})]),
\label{eq:13}
\end{equation}
where the probability distribution of $\mathbf{M}(\mathbf{r})$ is defined as 
\begin{equation}
P[\mathbf{M}(\mathbf{r})]=Ce^{-\beta H_{S}}e^{-\beta \Delta F},
\label{eq:14}
\end{equation}
and $C$ is a normalization constant. The functional integration in Eq.~(\ref
{eq:13}) expresses a summation of the contributions coming from all space
profiles of magnetization $\{\mathbf{M}(\mathbf{r})\}$ contributing to
given value of the exchange field $\mathbf{\Delta }$. This necessitates the
functional integration over all possible ''paths''\ of $\{\mathbf{M}(\mathbf{%
r})\},$ with the probability density $P[\mathbf{M}]$.\cite{3} Generalization
of Eqs.(\ref{eq:12})-(\ref{eq:13}) to $N$ -polaron case leads to the
following $N-$component Ginzburg-Landau Hamiltonian: 
\begin{multline}
H_{S}[\mathbf{M}_{1},...,\mathbf{M}_{N}]= \\
\sum_{i=1}^{N}\int d^{3}r_{i}(\frac{1}{2}\kappa ^{2}\sum_{j=1}^{3}\left|
\nabla _{i}\mathbf{M}_{ij}(\mathbf{r}_{i})\right| ^{2}+\frac{1}{2\chi }%
\left| \mathbf{M}_{i}(\mathbf{r}_{i})\right| ^{2}).
\end{multline}
Next, we define the probability distribution of exchange fields $\mathbf{%
\Delta }_{i}$, with $i=1,...,N,$ in the following form 
\begin{multline}
P(\mathbf{\Delta }_{1},...,\mathbf{\Delta }_{N})=  \label{eq:16} \\
C^{\prime }\int \prod_{i}^{N}DM_{i}e^{-\beta H_{S_{i}}}\ \delta (\mathbf{%
\Delta }_{i}-\mathbf{\Delta \lbrack M}_{i},w_{i}\mathbf{]})e^{-\beta \Delta
F_{T}}
\end{multline}
where $\Delta F_{T}$ is the electronic part of the total free energy of $N$
polarons system, $C^{\prime }$ is now the overall normalization factor, and 
\begin{equation}
\mathbf{\Delta \lbrack M}_{i},w_{i}\mathbf{]}=\frac{\alpha }{g\mu _{B}}\int 
\mathbf{M}_{i}(\mathbf{r})\left| w_{i}(\mathbf{r})\right| ^{2}d^{3}r.
\label{eq:17}
\end{equation}
Transformation made in Eq.~(\ref{eq:16}) defines our dynamic variables $%
\mathbf{\Delta }_{i}$ and the functional integrations in this equation can
be carried out for each $\mathbf{\Delta \lbrack M}_{i},w_{i}\mathbf{]}$ with
the methodology developed in Ref. 8.

When coupling between BMPs is important, our eigenvalues are given by Eq.~(%
\ref{34}) and the two-isolated BMP problem\ transforms now to the problem of
interacting polarons with $\mathbf{\Delta }^{+}$ and $\mathbf{\Delta }^{-}.$

\begin{figure}[tbp]
\includegraphics{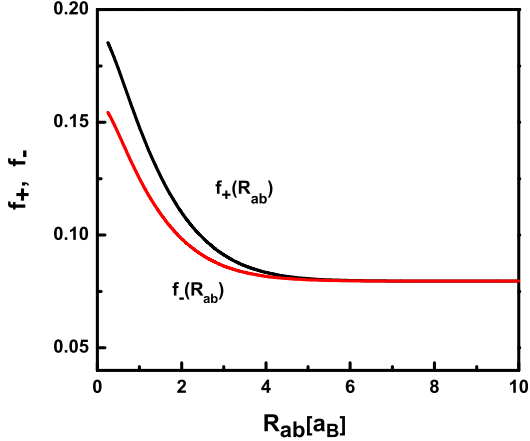} \label{fig4}
\caption{(Color online) Interpolaron distance dependence of the functions $%
f_{+}(R_{ab})$ and $f_{-}(R_{ab})$ describing the fluctuating exchange
field, as defined in Eq. (\ref{eq:v19}).}
\end{figure}

Thermodynamics of this system can be derived from Eq.~(\ref{eq:16}), which
after executing the functional integrations provides the physical free
energy: 
\begin{equation}
F=k_{B}T\ln \iint d^{3}\Delta ^{+}d^{3}\Delta ^{-}P(\mathbf{\Delta }^{+},%
\mathbf{\Delta }^{-}),  \label{eq:18}
\end{equation}
where the probability distribution of the exchange fields $\mathbf{\Delta }%
^{+},\mathbf{\Delta }^{-}$ is: 
\begin{multline}
P(\mathbf{\Delta }^{+},\mathbf{\Delta }^{-})=C^{\prime \prime }\exp \left\{ -%
\frac{(\mathbf{\Delta }^{+})^{2}}{8\varepsilon _{+}k_{B}T}\right\} \\
\exp \left\{ -\frac{(\mathbf{\Delta }^{-})^{2}}{8\varepsilon _{-}k_{B}T}%
\right\} \left\{ \sum_{i=1}^{6}\exp \left( -\frac{E_{i}}{k_{B}T}\right)
\right\} ,
\end{multline}
and the eigenvalues $E_{i}$ are the roots of Eq.~(\ref{30}), whereas the two
new parameters, $\varepsilon _{+}$ and .$\varepsilon _{-}$ are defined as
follows:

\begin{equation*}
\varepsilon _{\pm }(R_{ab})\equiv \frac{1}{4}\frac{\alpha ^{2}\chi }{(g\mu
_{B})^{2}}f_{\pm }(R_{ab})
\end{equation*}
\begin{equation}
\equiv \frac{1}{4}\frac{\alpha ^{2}\chi }{(g\mu _{B})^{2}}\int d^{3}r[\left|
w_{a}(r)\right| ^{2}\pm \left| w_{b}(r)\right| ^{2}]^{2}d^{3}r\,.
\label{eq:v19}
\end{equation}
Note that $P(\mathbf{\Delta }^{-},\mathbf{\Delta }^{+})$ has a correct
asymptotic behavior for the interpolaron distance $R_{ab}\rightarrow \infty
. $ We stress also, that the parameters $\varepsilon _{+}$ and $\varepsilon
_{-}$ are material dependent ($\chi $ is the spin system susceptibility)\
and correspond to the single-polaron parameter $\varepsilon _{p}$ of DS.\cite
{3} Their $R_{ab}$ dependence is crucial, as it affects the magnitudes of
either $\mathbf{\Delta }^{+}$ or $\mathbf{\Delta }^{-}$. In Fig. 4 we plot
this dependence. One sees, that for all finite distances $f_{+}\geq f_{-}$,
i.e. the field responsible for the triplet configurations of the two
impurity electrons is dominant. Note also, that even though the parameter $%
\beta $ $\rightarrow \infty $ in the limit $R_{ab}\rightarrow 0,$\ the
corresponding functions $f_{+}(R_{ab})$ and $f_{-}(R_{ab})$ are finite then
(see also below).

The dominating character of $\varepsilon _{+}$ may result in the
ferromagnetic-ground-state appearance\ of the BMPM. However, to prove that
explicitly we have to include also the effect of the exchange fields on the
eigenvalues $\left\{ E_{i}\right\} $. This can be done by determining the
most probable values of the fields $\mathbf{\Delta }^{+}$ and $\mathbf{%
\Delta }^{-}.$ For that purpose, we consider first the lowest-order solution
assuming that $E_{i}\thickapprox E_{0i}$. In analogy to the single-polaron
theory, we can now calculate the most probable values $\overline{\Delta ^{+}}
$ and $\overline{\Delta ^{-}}$ from the conditions: 
\begin{equation}
\left[ \partial P(\Delta ^{+},\Delta ^{-})/\partial \Delta ^{+}\right] _{%
\overline{\Delta ^{+}}}=\left[ \partial P(\Delta ^{+},\Delta ^{-})/\partial
\Delta ^{-}\right] _{\overline{\Delta ^{-}}}=0,  \label{eq:32}
\end{equation}
which lead to the following system of two coupled equations:%
\begin{widetext}
\begin{equation}
(1+A+B)\overline{\Delta ^{-}}^{3}-4\varepsilon _{-}\frac{\overline{\Delta
^{-}}^{3}}{\sqrt{\frac{1}{4}J^{2}+(\overline{\Delta ^{-}})^{2}}}C\tanh \left[
\frac{\sqrt{\frac{1}{4}J^{2}+(\overline{\Delta ^{-}})^{2}}}{k_{B}T}\right]
-8(1+A+B)\overline{\Delta ^{-}}\varepsilon _{-}k_{B}T=0,  \label{eq:33}
\end{equation}
and 
\begin{equation}
(B^{-1}+1+AB^{-1})\overline{\Delta ^{+}}^{3}-4\varepsilon _{+}\overline{\Delta ^{+}}^{2}\tanh \left[ \frac{\overline{\Delta ^{+}}}{k_{B}T}\right]
-8(1+B^{-1}+AB^{-1})\overline{\Delta ^{+}}\varepsilon _{+}k_{B}T=0,
\label{eq:34}
\end{equation}
where $A=A(T),$ $B=B(T)$ and $C(T)$ are defined as: 
\begin{equation}
A(T)\equiv \cosh \left[ \frac{\sqrt{\frac{1}{4}J4^{2}+(\overline{\Delta ^{-}})^{2}}}{k_{B}T}\right] \exp \left[ \frac{\frac{J6}{2}+L_{5}-\frac{J4}{2}-L}{k_{B}T}\right] \cosh ^{-1}\left[ \frac{\sqrt{\frac{1}{4}J^{2}+(\overline{\Delta ^{-}})^{2}}}{k_{B}T}\right] ,
\end{equation}
\begin{equation}
B(T)\equiv \cosh \left[ \frac{\overline{\Delta ^{+}}}{k_{B}T}\right] \exp \left[ \frac{\frac{J}{2}+L_{5}-L}{k_{B}T}\right] \cosh ^{-1}\left[ \frac{\sqrt{\frac{1}{4}J^{2}+(\overline{\Delta ^{-}})^{2}}}{k_{B}T}\right] ,
\end{equation}\end{widetext}%
%
and 
\begin{equation}
C(T)=1+\frac{\sqrt{\frac{1}{4}J^{2}+(\overline{\Delta ^{-}})^{2}}}{\sqrt{%
\frac{1}{4}J_{4}^{2}+(\overline{\Delta ^{-}})^{2}}}\frac{\sinh \left[ \frac{%
\sqrt{\frac{1}{4}J_{4}^{2}+(\overline{\Delta ^{-}})^{2}}}{k_{B}T}\right] }{%
\sinh \left[ \frac{\sqrt{\frac{1}{4}J^{2}+(\overline{\Delta ^{-}})^{2}}}{%
k_{B}T}\right] },  \label{edr}
\end{equation}
with $J=L_{6}-L_{5}$ and $J_{4}=L_{4}-L.$ This system of coupled equations
leads to the following three pairs of the solutions at $T=0,$ depending on
the sign of $\Delta E=E_{5}-E_{3}$: 
\begin{multline*}
(i)\qquad \overline{\Delta ^{-}}^{2}=(2\varepsilon _{-})^{2}-(J/2)^{2},\text{
and\quad }\overline{\Delta ^{+}}=2\varepsilon _{+}, \\
\text{for  }\Delta E=0,
\end{multline*}
\begin{multline}
(ii)\qquad \overline{\Delta ^{+}}=4\varepsilon _{+},\ \text{and\quad }%
\overline{\Delta ^{-}}=0,  \label{eq;v21} \\
\text{for  }\Delta E>0,
\end{multline}
and 
\begin{multline*}
(iii)\qquad \overline{\Delta ^{+}}=0\text{ and\quad }\overline{\Delta ^{-}}=%
\sqrt{(4\varepsilon _{-})^{2}-(J/2)^{2}}, \\
\text{for \ }\Delta E<0.
\end{multline*}
The condition $\Delta E(R_{ab})=0$ defines the critical distance $R_{c}$\
for $R_{ab}>R_{c}$\ the ground state is ferromagnetic. Therefore, the
driving interaction which aligns at $T=0$ individual polaron polarization
clouds can be defined at $R_{c}$\ in the form $\Delta E_{c}\equiv
E_{5}(iii)-E_{3}(ii).$ Let us then discuss explicitly the limit $%
T\rightarrow 0,$ for which the ground state is determined by the sign of the
expression: 
\begin{multline}
\left| \frac{J(R_{ab})}{2}+L_{5}-\sqrt{\frac{1}{4}J(R_{ab})^{2}+(\overline{%
\Delta ^{-}}(R_{ab}))^{2}}\right|  \label{eq:37} \\
-\left| L-\overline{\Delta ^{+}}(R_{ab})\right| \equiv f(R_{ab}).
\end{multline}
This expression for negative values yields the ground state belonging to
eigenvalue $E_{5},$ i.e. a mixture of two singlet states of impurity
electrons, whereas for $f(R_{ab})>0$ it leads to ferromagnetic, ($%
s_{tot}^{z}=\pm 1$) ground state with the corresponding eigenvalue $E_{3}$.
Therefore, the condition $f(R_{ab})=0$ defines a critical interpolaron
distance $R_{c}$, at which the crossover from magnetic to nonmagnetic ground
states occurs. A direct analysis shows that $\lim_{T\rightarrow 0}A(T)=0$
and $\lim_{T\rightarrow 0}C(T)=1,$ whereas an important role on the
character of the ground state is played by the function $B(T)$. This become
clearly visible if one determines its value for $T\rightarrow 0.$ Namely, $%
B(T\rightarrow 0)\rightarrow \infty $\ for $f(R_{ab})>0,$ so $%
B^{-1}(T\rightarrow 0)\rightarrow 0.$ It can be readily seen, that in this
case we have: 
\begin{equation}
\overline{\Delta ^{+}}=4\varepsilon _{+},\quad \text{and \ \ \ }\overline{%
\Delta ^{-}}=0.  \label{eq:38}
\end{equation}
In the other case, i.e. for $f(R_{ab})<0$, $B(T\rightarrow 0)\rightarrow 0$
and we find: 
\begin{equation}
\overline{\Delta ^{-}}=\sqrt{(4\varepsilon _{-})^{2}-(J(R_{ab})/2)^{2}}%
,\quad \text{and\quad }\overline{\Delta ^{+}}=0.\quad  \label{eq:39}
\end{equation}
Such peculiar behavior of\ $B(T\rightarrow 0)$ stabilizes the ground state
and moreover, makes our analytical solution exact, because $E_{i}=E_{0i}$
under these circumstances. For completeness, we need to determine still $%
B(T\rightarrow 0)$ at $R_{c}.$ It can be shown that $B(T\rightarrow
0)\rightarrow 1,$ with: 
\begin{equation}
\overline{\Delta ^{-}}=\sqrt{(2\varepsilon _{-})^{2}-(J(R_{c})/2)^{2}},\quad 
\text{and\quad }\overline{\Delta ^{+}}=2\varepsilon _{+}.  \label{eq:40}
\end{equation}
Even at low nonzero temperature solutions for $\overline{\Delta ^{+}}$ and $%
\overline{\Delta ^{-}}$ become complicated and temperature dependent.
Therefore, considering behavior of the state population probabilities at low 
$T,$\ one can propose an approximate expression for determining $R_{c}$: 
\begin{multline}
\frac{f_{+}-bf_{-}}{f_{-}}-\frac{L-bL_{5}}{2\varkappa f_{-}}+\frac{bJ}{%
4\varkappa f_{-}}  \label{eq:41} \\
-\frac{b(J/2)^{2}}{(4\varkappa f_{-})^{2}}\equiv w(R_{ab}),
\end{multline}
\ where $J_{H_{2}}(R_{ab})=$ $J(R_{ab})$ is taken for the hydrogen molecule
(i.e. $\varepsilon =1$ and $m^{\ast }/m_{e}=1,m_{e}$ being the electron
mass), whereas $R_{c}$ is expressed in units of $a_{B},$ and $\varkappa $\
is defined as: 
\begin{equation}
\varkappa \equiv \frac{\alpha ^{2}\chi }{4(g\mu _{B})^{2}}\frac{\varepsilon
^{2}m_{e}}{m^{\ast }}  \label{eq:42}
\end{equation}
and contains all\ the key material parameters that determine $R_{c},$
provided that the parameters $L,$ $L_{5},$ $J_{15}$ and $J_{56}$ are
calculated for the H$_{2}$ molecule. Note that neglecting the quadratic term
in the condition $w(R_{ab})=0$ and for $b=1$ is equivalent to $\Delta E=0.$
Next, we discuss the $R_{c}$ dependence on material parameters.

\begin{figure}[tbp]
\includegraphics{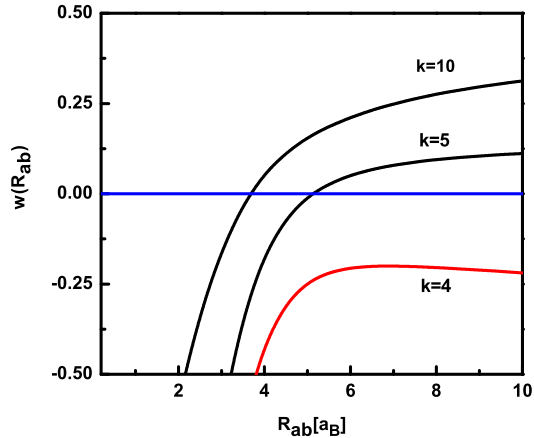} \label{fig5}
\caption{(Color online) Plot of $w(R_{ab}),$ \ defined in Eq. (59), for
three DMS characterized by $\varkappa =k\varkappa _{0}$, where the value $%
\varkappa _{0}$ corresponds to that for Cd$_{0.95}$Mn$_{0.05}$Se and $k>1$.}
\end{figure}

In Fig. 5 we plot $w(R_{ab})$ with $b=1.1$ ($T\simeq 1.1K$) for DMS
characterized by $\varkappa =k\varkappa _{0}$, with $k=4,5,$ and $15$, where
the value $\varkappa _{0}$ corresponds to that for Cd$_{0.95}$Mn$_{0.05}$Se (%
$\varepsilon =9.4,$ $m^{\ast }/m_{e}=0.13$, ($\overline{\Delta }%
=2\varepsilon _{p}\lesssim 1$ meV)).\cite{3} One sees from Fig.2 that for Cd$%
_{0.95}$Mn$_{0.05}$Se, $w(R_{ab})<0$ for all $R_{ab};$ then the $%
s_{tot}^{z}=0$ spin configuration of the impurity electrons is always stable.

So far we have treated the magnetic susceptibility $\chi $ of localized
magnetic ions as a material parameter neglecting its strong temperature
dependence. In Cd$_{0.95}$Mn$_{0.05}$Se and at low temperature, $\chi $
takes the forme of Curie-Weiss law $\chi =C_{M}/(T+T_{0})$ with $T_{0}=1.2$
K. Properties of the system at low temperatures may be deduced from analysis
of the averages of $\Delta ^{+}$ and $\Delta ^{-},$ as they reflect
properties of the most probable values $\overline{\Delta }^{+}$ and $%
\overline{\Delta }^{-}$ expressed via conditions $(i)-(iii)$.\ In order to
see that within our approach the ferromagnetic configuration of the impurity
electron is possible in general, we have displayed in Fig. 5 the numerical
average of $\Delta ^{+}$ and $\Delta ^{-}$ vs. $T$ for exemplary value of
the parameter $\varkappa (T)=k\varkappa _{0}(T)$, with $k=25$. 
\begin{figure}[tbp]
\includegraphics{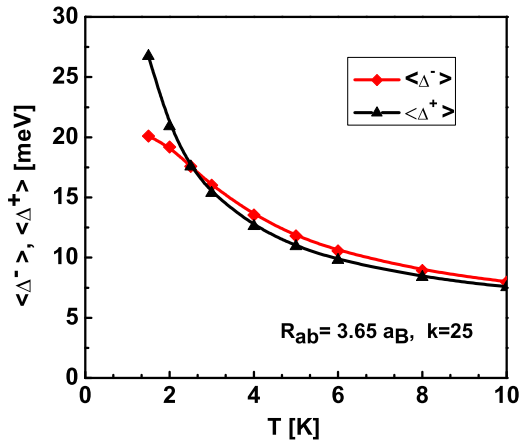} \label{fig6}
\caption{(Color online) Temperature dependence of average exchange fields $%
\Delta ^{+}$ and $\Delta ^{-}$ for DMS characterized by $\varkappa
=25\varkappa _{0}$.}
\end{figure}
Such value of $k$\ for n-type DMS is naturally outside of accessible range
of material parameter, but for the p-type DMS the p-d exchange is about $4-5$
times stronger then that for the n-type, then such high value of $k$ is
justified. For this value of $k$\ one finds $\Delta E_{c}\approx 0.47$ meV
(calculated at $2$ K,  with $T_{0}=1.2$ K and for $R_{ab}=3.65$ a$_{B}$), which allow for a very
rough estimation of the Curie temperature $T_{C}$, in the mean field
approximation for a 3D cubic BMP lattice, would be $22$ K.

For completeness, we plot in Fig. 6 temperature dependence of the average
values of $\mathbf{\Delta }^{+}$ and $\mathbf{\Delta }^{-}$ taking the
parameters for Cd$_{0.95}$Mn$_{0.05}$Se at $R_{ab}=2$ and $4.25$ a$_{B}$. As
can see, the numerical results confirm that for this material a nonmagnetic (%
$s_{tot}^{z}=0$) configuration of the impurity electrons is always stable.

\section{Conclusions}

In this work we have developed the model of BMP molecule (BMPM) consisting
of two overlapping polarons, which have arbitrarily oriented in space their
polaronic spin clouds of arbitrary magnitude. Their mutual interaction is
accounted for within the molecular electronic states. We also succeeded in
deriving thermodynamics of these states that allows to analyse their
properties at nonzero temperature. These features allow for a better
understanding of the earlier results, including the range of their
applicability. More directly, an important role of an accurate inclusion of the
impurity electron wave functions has been demonstrated. What is also
important, the microscopic interpretation of Wolff-Bhatt-Durst model
Hamiltonian-parameters\cite{12} has been provided. Simultaneously, the
regular case of the many-level generalized Hubbard model with random fields
of Angelescu-Bhatt\cite{11} of nonoverlapping, large polarons has been
extended to the case of overlapping polarons forming the BMP molecule.

Generally, our model of BMPM in DMS confirms a possibility of the ferromagnetic
ground-state appearance for certain DMS materials and appropriate distances between
polarons. One of the important and new results of our approach is a direct
incorporation of material parameters into the BMP molecule model. Presented
here numerical calculation for BMP molecule embodied in Cd$_{0.95}$Mn$_{0.05}
$Se shows, that for all interpolaron distances the polaronic molecule ground
state spin configuration is $s_{tot}^{z}=0$ for the impurity electrons.
Then, in this DMS a non-magnetic ground state of BMP molecule is always
stable. From the other side, our model predicts a ferromagnetic
(spin-triplet) ground state of BMP molecule for material parameters
corresponding to p-type DMSs. The nature of the ground state
depends on magnitude of the parameter $\varkappa $, defined by Eq.~(\ref
{eq:42}), which is proportional to the magnetic susceptibility of the spin
system of host DMS. Therefore, the RKKY interaction responsible for the low temperature ferromagnetic
ordering in p-type DMSs enhances ferromagnetic ground state
of BMPM when approaching the critical region from the high-temperature side.
However, the results for the p-type DMSs must be analyzed within the approach, which accounts
correctly for the non-s-type character of BMP carrier wave function, as well as non-Gaussian contributions to the free energy functional. Such formulation of BMPM is under consideration and will be presented separately.

\appendix

\section{The matrix representation of $H_{I}$ within first
quantization formalism}

We assume that the exchange field acting on impurity electrons is relatively
weak and that orbital moments can be disregarded. As a consequence, the
two-impurity electron wave function may by written as a product of
functions, which depend on the spatial and the spin variables separately
i.e. it is of the form 
\begin{equation}
\Psi _{c}(\left\{ \widehat{\mathbf{S}}_{i}\right\} ;\mathbf{r}_{1},\mathbf{r}%
_{2},\sigma _{1},\sigma _{2})=\chi _{S}\left\{ \widehat{\mathbf{S}}%
_{i}\right\} \chi _{\sigma 1\sigma 2}\Psi _{c}(\mathbf{r}_{1},\mathbf{r}%
_{2})\,,  \tag{A1}  \label{eq:v2}
\end{equation}
where $\chi \left\{ \widehat{\mathbf{S}}_{i}\right\} $ is the wave function
of the spins, and $\chi _{\sigma 1\sigma 2}$\ is the spin part of that for
the carriers, whose spatial wave function is $\Psi _{c}(\mathbf{r}_{1},%
\mathbf{r}_{2}),$ where $c$ runs over all pairs ($ab,aa,bb$). We take the
two-electron functions as composed of orthogonal molecular wave functions $%
w_{a,b}(r_{i})$: 
\begin{equation}
w_{a,b}(r)=\beta \lbrack \psi _{a,b}(r)-\gamma \psi _{b,a}(r)],  \tag{A2}
\label{eq:v3}
\end{equation}
contain 1s-hydrogenic-like wave functions $\psi _{a}$ and $\psi _{b},$ with
the effective Bohr radius a$_{B},$with $\beta $\ and $\gamma $ being the
mixing coefficients. These orthogonal single-particle wave functions are
used to construct the two-particle six-dimensional space. The space part is
identical with that for the hydrogeniclike molecule. Therefore, we can write
down directly the following six two-electron wave functions: 
\begin{equation*}
\Psi _{1,4}\chi _{1,4}=\frac{1}{\sqrt{2}}[w_{a}(\mathbf{r}_{1})w_{b}(\mathbf{%
r}_{2})\mp w_{a}(\mathbf{r}_{2})w_{b}(\mathbf{r}_{1})]
\end{equation*}
\begin{equation}
\times \frac{1}{\sqrt{2}}[\chi _{1/2}(1)\chi _{-1/2}(2)\pm \chi
_{-1/2}(1)\chi _{1/2}(2)]\,,  \tag{A3}  \label{eq:v4}
\end{equation}
for the spin-triplet state with $s_{tot}^{z}=0$ and the singlet state
respectively and 
\begin{equation*}
\Psi _{2,3}\chi _{2,3}=\frac{1}{\sqrt{2}}[w_{a}(\mathbf{r}_{1})w_{b}(\mathbf{%
r}_{2})-w_{a}(\mathbf{r}_{2})w_{b}(\mathbf{r}_{1})]
\end{equation*}
\begin{equation}
\times \lbrack \chi _{\pm 1/2}(1)\chi _{\pm 1/2}(2)]\,,  \tag{A4}
\label{eq:v5}
\end{equation}
for the remaining triplet states with $s_{tot}^{z}=\pm 1,$ and 
\begin{equation*}
\Psi _{5,6}\chi _{5,6}=\frac{1}{\sqrt{2}}[w_{a}(\mathbf{r}_{1})w_{a}(\mathbf{%
r}_{2})\mp w_{b}(\mathbf{r}_{2})w_{b}(\mathbf{r}_{1})]
\end{equation*}
\begin{equation}
\times \frac{1}{\sqrt{2}}[\chi _{1/2}(1)\chi _{-1/2}(2)\pm \chi
_{-1/2}(1)\chi _{1/2}(2)],  \tag{A5}  \label{eq:v6}
\end{equation}
for the two ionic singlet states.

Next, we find the matrix representation of $H_{I}$\ by averaging over the
spatial coordinates, but keeping out the carrier spin \ still in the
operator form: 
\begin{equation*}
\left\langle \Psi _{n}\left| H_{I}\right| \Psi _{m}\right\rangle =\frac{%
\alpha }{g\mu _{B}}[\widehat{\mathbf{s}}_{1}\cdot \int \Psi _{n}^{\ast }%
\mathbf{M}(\mathbf{r}_{1})\Psi _{m}d^{3}r_{1}d^{3}r_{2}
\end{equation*}
\begin{equation}
+\widehat{\mathbf{s}}_{2}\cdot \int \Psi _{n}^{\ast }\mathbf{M}(\mathbf{r}%
_{2})\Psi _{m}d^{3}r_{1}d^{3}r_{2}]\,.  \tag{A6}  \label{eq:v7}
\end{equation}
Taking into account the symmetry properties of spatial wave function with
respect to the transposition of the carrier coordinates, the matrix elements
of $H_{I}$ can be recast to the form: 
\begin{equation*}
\left\langle \Psi _{n}\left| H\right| \Psi _{m}\right\rangle =\widehat{%
\mathbf{s}}_{1}\cdot \mathbf{\Delta }_{1}[\mathbf{M,}\Psi _{n},\Psi _{m}]+%
\widehat{\mathbf{s}}_{2}\cdot \mathbf{\Delta }_{2}[\mathbf{M,}\Psi _{n},\Psi
_{m}]
\end{equation*}
\begin{equation}
\equiv (\widehat{\mathbf{s}}_{1}+sign(\Psi _{n})sign(\Psi _{m})\widehat{%
\mathbf{s}}_{2})\cdot \mathbf{\Delta }[\mathbf{M,}\Psi _{n},\Psi _{m}]\,, 
\tag{A7}  \label{eq:v8}
\end{equation}
where the factor $sign(...)=\pm 1$ and\ expresses the parity of the
corresponding wave function. The exchange fields, $\mathbf{\Delta }_{l}[%
\mathbf{M,}\Psi _{n},\Psi _{m}]$ are thus defined as: 
\begin{equation}
\mathbf{\Delta }_{l}[\mathbf{M,}\Psi _{n},\Psi _{m}]\equiv \frac{\alpha }{%
g\mu _{B}}\int \int \Psi _{n}^{\ast }\mathbf{M}(\mathbf{r}_{l})\Psi
_{m}d^{3}r_{1}d^{3}r_{2}\,,  \tag{A8}  \label{eq:v9}
\end{equation}
Note that the fields $\mathbf{\Delta }_{l}[\mathbf{M,}\Psi _{n},\Psi _{m}]$
are equal to $\mathbf{\Delta }[\mathbf{M,}\Psi _{n},\Psi _{m}],$ except for
the sign. An explicit calculation provides a remarkable reduction of the
number of the exchange fields. Namely, only the following two appear in the
final BMP pair Hamiltonian matrix: 
\begin{equation}
\mathbf{\Delta }^{\pm }[\mathbf{M}]=\frac{\alpha }{g\mu _{B}}\int
d^{3}r[\left| w_{a}(r)\right| ^{2}\pm \left| w_{b}(r)\right| ^{2}]\mathbf{M}%
(r)\,.  \tag{A9}  \label{eq:v10}
\end{equation}
The labeling of the exchange fields $\mathbf{\Delta }^{+}$and $\mathbf{%
\Delta }^{-}$ originates from the wave function parities (e.g. ''+''
corresponds to the triplet-triplet matrix elements and ''-'' to the
singlet-triplet ones).

Here we outline the details of calculations of the matrix elements $%
\left\langle \Phi _{\tau }^{s_{tot},s_{tot}^{z}}\left| H_{BMP}\right| \Phi
_{\nu }^{s_{tot}^{\prime },s_{tot}^{z\prime }}\right\rangle $ in the
singlet-triplet basis. According to the definition given in the text : $\Phi
_{\tau }^{s,s_{tot}^{z}}=\Psi _{\tau }\chi _{\tau }^{s_{tot},s_{tot}^{z}},$
and taking into account that Hamiltonian $H_{II}$ does not contain
explicitly the carrier spin variables, one can write: 
\begin{multline}
\left\langle \Phi _{\tau }^{s_{tot},s_{tot}^{z}}\left| H_{BMP}\right| \Phi
_{\nu }^{s_{tot}^{\prime },s_{tot}^{z\prime }}\right\rangle   \tag{A10} \\
=\left\langle \chi _{\tau }^{s_{tot},s_{tot}^{z}}\left| \left\langle \Psi
_{\tau }\left| H_{BMP}\right| \Psi _{\nu }\right\rangle \right| \chi _{\nu
}^{s_{tot}^{\prime },s_{tot}^{z\prime }}\right\rangle  \\
=E_{\tau }\delta _{\tau \nu }+\left\langle \chi _{\tau
}^{s_{tot},s_{tot}^{z}}\left| (\widehat{\mathbf{s}}_{1}+sign(\tau \nu )%
\widehat{\mathbf{s}}_{2})\cdot \mathbf{\Delta }^{\tau \nu }]\right| \chi
_{\nu }^{s_{tot}^{\prime },s_{tot}^{z\prime }}\right\rangle ,
\end{multline}
where $\widehat{\mathbf{s}}_{1}=\widehat{\mathbf{s}}_{1}\otimes \mathbf{1}$, 
$\widehat{\mathbf{s}}_{2}=\mathbf{1}\otimes \widehat{\mathbf{s}}_{2}$ are
the carrier spin operators and the corresponding spinors $\chi _{\tau
}^{s_{tot},s_{tot}^{z}}$ are explicitly defined below: 
\begin{equation}
\chi _{+}^{0,0}=\frac{1}{\sqrt{2}}[\chi _{1/2}(1)\otimes \chi
_{-1/2}(2)-\chi _{-1/2}(1)\otimes \chi _{1/2}(2)],  \tag{A11}
\end{equation}
\begin{equation}
\chi _{-}^{1,0}=\frac{1}{\sqrt{2}}[\chi _{1/2}(1)\otimes \chi
_{-1/2}(2)+\chi _{-1/2}(1)\otimes \chi _{1/2}(2)],  \tag{A12}
\end{equation}
\begin{equation}
\chi _{-}^{1,1}=\chi _{1/2}(1)\otimes \chi _{1/2}(2),  \tag{A13}
\end{equation}
and 
\begin{equation}
\chi _{-}^{1,-1}=\chi _{-1/2}(1)\otimes \chi _{-1/2}(2).  \tag{A14}
\end{equation}
To short-hand notation of the tensor product symbol are omitted below. We
recall that we have fixed the direction of the spin quantization axis as
parallel to the local exchange field $\mathbf{\Delta }^{-}[\mathbf{M}]$.

Generally, one can assume that in the selected coordinate system direction
of $\mathbf{\Delta }^{+}[\mathbf{M}]$ is described by the angles $\theta $
and $\phi $. Therefore, evaluation of Eq. (A10) with these exchange fields
is the simplest in the appropriately rotated spin basis, namely: 
\begin{equation}
\chi _{1/2}=\chi _{\theta \phi }\func{e}^{i\frac{\phi }{2}}\cos \frac{\theta 
}{2}-\chi _{\overline{\theta }\overline{\phi }}\func{e}^{i\frac{\phi }{2}%
}\sin \frac{\theta }{2},  \tag{A15}
\end{equation}
\begin{equation}
\chi _{-1/2}=\chi _{\theta \phi }\func{e}^{-i\frac{\phi }{2}}\sin \frac{%
\theta }{2}+\chi _{\overline{\theta }\overline{\phi }}\func{e}^{-i\frac{\phi 
}{2}}\cos \frac{\theta }{2},  \tag{A16}
\end{equation}
where the angles $\overline{\theta }$ and$\overline{\text{ }\phi }$ describe
the spin state with an arbitrary direction opposite to that defined by $%
\theta $ and $\phi .$

From Eq. (A10), the diagonal matrix elements can be written as follows

\begin{multline}
\left\langle \chi _{\tau }^{s_{tot},s_{tot}^{z}}\left| H_{BMP}\right| \chi
_{\nu }^{s_{tot}^{\prime },s_{tot}^{z\prime }}\right\rangle  \tag{A17} \\
=E_{\tau }+\left\langle \chi _{\tau }^{s_{tot},s_{tot}^{z}}\left| (\widehat{%
\mathbf{s}}_{1}+\widehat{\mathbf{s}}_{2})\cdot \mathbf{\Delta }^{\tau
}\right| \chi _{\tau }^{s_{tot},s_{tot}^{z}}\right\rangle .
\end{multline}

Note that $\left\langle \chi _{+}^{0,0}\left| H_{BMP}\right| \chi
_{+}^{0,0}\right\rangle =E_{+}$ because $\chi _{+}^{0,0}$ transform as a
scalar under rotations, i.e. 
\begin{multline}
\chi _{1/2}(1)\chi _{-1/2}(2)-\chi _{-1/2}(1)\chi _{1/2}(2)  \tag{A18} \\
=\chi _{\theta \phi }(1)\chi _{\overline{\xi }\overline{\phi }}(2)-\chi _{%
\overline{\theta }\overline{\phi }}(1)\chi _{\xi \phi }(2).
\end{multline}
Also, the matrix element $\left\langle \chi _{-}^{1,0}\left| H_{BMP}\right|
\chi _{-}^{1,0}\right\rangle $ can be computed in the following manner 
\begin{widetext}
\begin{multline}
\left\langle \chi _{-}^{1,0}\left\vert (\widehat{\mathbf{s}}_{1}+\widehat{\mathbf{s}}_{2})\cdot \mathbf{\Delta }^{--}\right\vert \chi
_{-}^{1,0}\right\rangle   \tag{A19} \\
=\frac{1}{2}\sum_{i}\left\langle \chi _{1/2}(1)\chi _{-1/2}(2)+\chi
_{-1/2}(1)\chi _{1/2}(2)\left\vert \widehat{\mathbf{s}}_{i}\cdot \mathbf{\Delta }^{--}\right\vert \chi _{1/2}(1)\chi _{-1/2}(2)+\chi _{-1/2}(1)\chi
_{1/2}(2)\right\rangle  \\
=\frac{1}{2}\sum_{i}\sum_{\sigma ,\sigma ^{\prime }}^{1/2,-1/2}\left\langle
\chi _{\sigma }(1)\chi _{\sigma ^{\prime }}(2)\left\vert \widehat{\mathbf{s}}_{i}\cdot \mathbf{\Delta }^{--}\right\vert \chi _{\sigma }(1)\chi _{\sigma
^{\prime }}(2)\right\rangle =\frac{1}{2}\sum_{i}\sum_{\sigma
}^{1/2,-1/2}\left\langle \chi _{\sigma }(i)\left\vert \widehat{\mathbf{s}}_{i}\cdot \mathbf{\Delta }^{--}\right\vert \chi _{\sigma }(i)\right\rangle 
\\
=\frac{1}{2}\sum_{i}^{1,2}\sum_{\sigma }^{\xi _{2}\phi _{2},\overline{\xi }_{2},\overline{\phi }_{2}}\left\langle \chi _{\sigma }(i)\left\vert
s_{i}^{z^{\prime }}\right\vert \chi _{\sigma }(i)\right\rangle \Delta ^{--}=\frac{1}{4}\Delta ^{--}\left[ \left( \cos ^{2}(\xi )-\sin ^{2}(\xi )\right)
+\left( \sin ^{2}(\xi )-\cos ^{2}(\xi )\right) \right] =0,
\end{multline}
\end{widetext}%
%
and thus $\left\langle \chi _{-}^{1,0}\left| H_{BMP}\right| \chi
_{-}^{1,0}\right\rangle =E_{-}.$ 

The other matrix elements are:
\begin{multline}
\left\langle \chi _{-}^{1,1}\left| H_{BMP}\right| \chi
_{-}^{1,1}\right\rangle =  \tag{A20} \\
\sum_{i}\left\langle \chi _{1/2}(i)\left| \widehat{\mathbf{s}}_{i}\cdot 
\mathbf{\Delta }^{+}\right| \chi _{1/2}(i)\right\rangle =\Delta ^{+}\cos
\theta ,
\end{multline}
and
\begin{multline}
\left\langle \chi _{-}^{1,-1}\left| H_{BMP}\right| \chi
_{-}^{1,-1}\right\rangle =  \tag{A21} \\
\sum_{i}\left\langle \chi _{-1/2}(i)\left| \widehat{\mathbf{s}}_{i}\cdot 
\mathbf{\Delta }^{+}\right| \chi _{-1/2}(i)\right\rangle =-\Delta ^{+}\cos
\theta .
\end{multline}

Finally, the non vanishing off-diagonal matrix elements are: 
\begin{multline}
\left\langle \chi _{+}^{0,0}\left| H_{BMP}\right| \chi
_{-}^{1,0}\right\rangle =  \tag{A22} \\
\Delta ^{-}\left\langle \chi _{+}^{0,0}\left| s_{1}^{z}-s_{2}^{z}\right|
\chi _{-}^{1,0}\right\rangle =\Delta ^{-}\left\langle \chi _{+}^{0,0}|\chi
_{+}^{0,0}\right\rangle ,
\end{multline}
\begin{multline}
\left\langle \chi _{-}^{1,0}\left| H_{BMP}\right| \chi
_{-}^{1,1}\right\rangle =\left\langle \chi _{-}^{1,0}\left| (\widehat{%
\mathbf{s}}_{1}+\widehat{\mathbf{s}}_{2})\cdot \mathbf{\Delta }^{+}\right|
\chi _{-}^{1,1}\right\rangle   \tag{A23} \\
=\frac{1}{\sqrt{2}}\sum_{j}\left\langle \chi _{-1/2}(j)\left| \widehat{%
\mathbf{s}}_{j}\cdot \mathbf{\Delta }^{+}\right| \chi _{1/2}(j)\right\rangle 
\notag \\
=\frac{1}{\sqrt{2}}\Delta ^{+}e^{i\varphi }\sin \theta ,  \notag
\end{multline}
and 
\begin{equation}
\left\langle \chi _{-}^{1,0}\left| H_{BMP}\right| \chi
_{-}^{1,-1}\right\rangle =\left\langle \chi _{-}^{1,0}\left| H_{BMP}\right|
\chi _{-}^{1,1}\right\rangle ^{\ast }.  \tag{A24}
\end{equation}
\bigskip 

Next, we select a convenient direction of the quantization axis for the
BMP-pair problem, which is selected as aligned with the field $\mathbf{%
\Delta }^{-}$. In effect, the matrix representation of $H_{I}$ in the six
dimensional basis of the two-electron wave functions is: 
\begin{equation}
\left( 
\begin{array}{cccccc}
0 & \Delta ^{+}\frac{\sin \theta }{\sqrt{2}}e^{i\varphi } & \Delta ^{+}\frac{%
\sin \theta }{\sqrt{2}}e^{-i\varphi } & \Delta ^{-} & 0 & 0 \\ 
\Delta ^{+}\frac{\sin \theta }{\sqrt{2}}e^{-i\varphi } & \Delta ^{+}\cos
\theta  & 0 & 0 & 0 & 0 \\ 
\Delta ^{+}\frac{\sin \theta }{\sqrt{2}}e^{i\varphi } & 0 & -\Delta ^{+}\cos
\theta  & 0 & 0 & 0 \\ 
\Delta ^{-} & 0 & 0 & 0 & 0 & 0 \\ 
0 & 0 & 0 & 0 & 0 & \Delta ^{-} \\ 
0 & 0 & 0 & 0 & \Delta ^{-} & 0
\end{array}
\right) \,,  \tag{A25}  \label{eq:v11}
\end{equation}
where, $\varphi $ and $\theta $ are respectively, the azimuthal and the
polar angles between the exchange field $\mathbf{\Delta }^{+}$ and $\mathbf{%
\Delta }^{-}$. This form of the Hamiltonian matrix is identical to that
taken in Sec. II for a detailed analysis.

\section{Equivalence of Eq. (35) with the solution for two isolated BMPs}

In this Appendix we demonstrate that our solution of the asymptotic case for 
$R_{ab}\longrightarrow \infty $ is equivalent to that know for two isolated
BMP, hence we prove the equivalency between Eq. (37) and Eq. (38). To do
this we need calculate first: 
\begin{equation}
2\Delta ^{+}\Delta ^{-}\cos (\theta )=(\mathbf{\Delta }^{+}+\mathbf{\Delta }%
^{-})^{2}-(\mathbf{\Delta }^{+}\mathbf{)}^{2}-(\mathbf{\Delta }^{-}\mathbf{)}%
^{2}.  \tag{B1}
\end{equation}
Taking now into account that: 
\begin{equation}
(\mathbf{\Delta }^{+}+\mathbf{\Delta }_{-})^{2}=4\mathbf{\Delta }_{a}^{2}, 
\tag{B2}
\end{equation}
and 
\begin{equation}
(\mathbf{\Delta }^{+}\mathbf{)}^{2}+(\mathbf{\Delta }^{-}\mathbf{)}^{2}=2(%
\mathbf{\Delta }_{a}^{2}+\mathbf{\Delta }_{b}^{2}),  \tag{B3}
\end{equation}
one can find: 
\begin{equation}
\mathbf{\Delta }^{+}\Delta ^{-}\cos (\theta )=\mathbf{\Delta }_{a}^{2}-%
\mathbf{\Delta }_{b}^{2}.  \tag{B4}
\end{equation}
Substitution of Eqs.:(B3)-(B4) into Eq.(37) gets finally: 
\begin{equation}
\left( 
\begin{array}{c}
2E_{\infty }+\frac{1}{2}(\Delta _{a}+\Delta _{b}) \\ 
2E_{\infty }-\frac{1}{2}(\Delta _{a}+\Delta _{b}) \\ 
2E_{\infty }+\frac{1}{2}(\Delta _{a}-\Delta _{b}) \\ 
2E_{\infty }-\frac{1}{2}(\Delta _{a}-\Delta _{b})
\end{array}
\right)   \tag{B5}
\end{equation}
These eigenvalues can be found also taking independent linear combination of
the solutions for the two isolated polarons. This constitutes the
equivalence between the two solutions.


\begin{thebibliography}{99}
\bibitem{1}  T. Jungwirth, J. Sinowa, J. Masek, J. Kucera, and A. H.
MacDonald, Rev. Mod. Phys. \textbf{78}, 809 (2006); T. Dietl et. al.,
Science \textbf{287} 1019 (2000).

\bibitem{Teresa}  J. M. De Teresa et al., Nature \textbf{386,} 256 (1997).

\bibitem{Coey}  J. M. Coey, M. Venkatesan and C. B. Fitzgerald, Nature
Materials \textbf{4}, 173 (2005).

\bibitem{Flor1}  K. Kikoin and V. Fleurov, Phys. Rev. B \textbf{74} 174407
(2006).

\bibitem{Flor2}  P.M. Krstaji\'{c} et al., Phys. Rev. B \textbf{70}, 195215
(2004).

\bibitem{Buzar}  R. Bouzerar, G. Bouzerar, and T. Ziman, Phys. Rev. B 
\textbf{73}, 024411 (2006); G. Bouzerar, T. Ziman, and J. Kudrnovsky, Phys.
Rev. B \textbf{72}, 125207 (2005).

\bibitem{2}  J. Spa\l ek, A. Lewicki, Z. Tarnawski, J.K. Furdyna, R.R. Ga\l
azka, and Z. Obuszko, Phys.Rev. B \textbf{33}, 3407 (1986).

\bibitem{3}  T. Dietl and J. Spalek, Phys. Rev. Lett. \textbf{48}, 355
(1982); Phys. Rev. B \textbf{28}, 1548 (1983). refered to as DS.

\bibitem{4}  S.M. Rybachenko and Y.G. Semenov, Zh. Eksp. Teor. Fiz. \textbf{%
84}, 1419 (1983). [Sov. Phys. JETP \textbf{57}, 825 (1983)].

\bibitem{5}  D. Heiman, P.A. Wolff and J. Warnock, Phys. Rev. B \textbf{27},
4848 (1983); P. A. Wolff and J. Warnock, J. Appl. Phys. \textbf{55}, 2300
(1984).

\bibitem{6}  A. Golnik, J. Ginter, and J.A. Gaj, J. Phys. C \textbf{16},
6073 (1983).

\bibitem{7}  T. H. Nhung, R. Planel, C. Benoit a la Guillaume, and A.K.
Bhattacharjee, Phys. Rev. B. \textbf{31}, 2388 (1985).

\bibitem{8}  A. Golnik and J. Spa\l ek, J. Magn. Magn. Mat. \textbf{54-57},
1207 (1986).

\bibitem{9}  M. Umehara, Phys. Rev. B \textbf{61}, 12209 (2000).

\bibitem{10}  P.A. Wolff, R.N. Bhatt, and A.C. Durst, J. Appl. Phys. \textbf{%
79}, 5196 (1996).

\bibitem{11}  D.E. Angelescu and R. N. Bhatt, Phys. Rev. B \textbf{65},
75211 (2002).

\bibitem{12}  A.C. Durst, R.N. Bhatt, and P.A. Wolff, Phys. Rev. B \textbf{65%
}, 235205 (2002).

\bibitem{13}  P. W. Anderson, Phys. Rev. B \textbf{115}, 2 (1959); K. A.
Chao, J. Spa\l ek, and A. M. Ole\'{s}, J. Phys. C \textbf{10}, L271 (1977).

\bibitem{in1}  M. Berciu and R.N. Bhatt, Phys. Rev. Lett. \textbf{90},
029702 (2003); M. Berciu and R.N. Bhatt, B \textbf{69}, 045202 (2004).

\bibitem{in2}  M.P. Kennett, M. Berciu, and R.N. Bhatt, Phys. Rev. B \textbf{%
66 }045207 (2002); M.P. Kennett, M. Berciu, and R.N. Bhatt, Phys. Rev. B 
\textbf{65} 115308 (2002); M. Berciu and R.N. Bhatt, ibid. \textbf{66},
085207 (2002).

\bibitem{in3}  L. Brey and G. Gomez-Santos, Phys. Rev. B \textbf{68}, 115206
(2003).

\bibitem{in4}  G. A. Fiete, G. Zarand, K. Damle, and C. P. Moca, Phys. Rev.
B \textbf{72} 045212 (2005); G. A. Fiete, G. Zarand and K. Damle, Phys. Rev.
Lett. \textbf{91} 097202 (2003).

\bibitem{in5}  A.Kaminski and S. Das Sarma, Phys. Rev. Lett. 88 247202
(2002); J. Priour and S. Das Sarma, Phys. Rev. Lett. \textbf{97}, 127201
(2006).

\bibitem{14}  see e.g. J. Spa\l ek, R. Podsiad\l y, W. W\'{o}jcik and A.
Rycerz, Phys. Rev. B \textbf{6},\textbf{1} 15676 (2000).

\end{thebibliography}
\end{document}